# A compressive sensing based parameter estimation for free-space continuous-variable quantum key distribution

Xiaowen Liu, Chen Dong, Xingyu Wang, and Tianyi Wu

*Abstract*—In satellite-based free-space continuous-variable QKD (CV-QKD), the parameter estimation for the atmospheric channel fluctuations due to the turbulence effects and attenuation is crucial for analyzing and improving the protocol performance. In this paper, compressive sensing (CS) theory is applied to free-space CV-QKD to achieve the channel parameter estimation with low computational complexity and small amount of data. According to CS theory, the possibility of the sparse representation for free-space channel is analyzed and the two types of sparse reconstruction models for the channel parameters are constructed combining with the stability of the sub-channels. The most part of variable for parameter estimation is saved by using the model relying on the variables in the quantum signals, while all the variables can be used to generate the secret key by using the model relying on the second-order statistics of the variables. The methods are well adapted for the cases with the limited communication time since a little or no variable is sacrificed for parameter estimation. Finally, simulation results are given to verify the effectiveness of the proposed methods.

## I. INTRODUCTION

Quantum key distribution (QKD) which is one of the most practical application of quantum communication is capable of applying the quantum channel to share the secure key between two distant partners conventionally called Alice and Bob, even though the quantum information may be eavesdropped by Eve [1-3]. Heisenberg's uncertainty principle and the quantum no-cloning theorem as the laws of quantum mechanics form the theoretical basis for the information theoretic-security regarding the QKD protocols [4,5]. Thereinto, continuous-variable QKD (CV-QKD) protocols where the information is encoded into the quadratures of phase space adopt the high-efficiency homodyne or heterodyne detection for the random coherent states sent by Alice, and have the advantage of higher channel capacity, easier to implement and more robust against noise than discrete-variable QKD (DV-QKD), etc [6-10]. Moreover, free-space CV-QKD which adopts the atmospheric channel to propagate the quantum states has potential to achieve long-distance communication due to low-loss transmission in outer space, even to facilitate the global quantum communication network establishment, and has more flexible to contact the mobile node than the fiber-optical infrastructure[11-13].

However, the atmospheric channel has the characteristics of fluctuation and time varying due to the turbulence effects and attenuation, and the transmittance fluctuations caused by turbulence effects such as beam wandering and scintillation usually introduce the additional excess noise to be destructive to free-space CV-QKD protocols, especially to the variance of the quadratures in the quantum signal [14-17]. Therefore, it is vital for free-space CV-QKD protocols to estimate the channel parameters (i.e. the transmittance and excess noise fluctuations) to assist the practical security analysis of the protocols, the post-processing procedure, and the channel fluctuation compensation, etc [18,19].

Currently, the methods for the estimation of the channel parameters fall into two major categories: atmospheric turbulence modelling and stabilization sub-channel parameter estimation. The elliptical beam model [20] is an early method for atmospheric turbulence analysis and is capable of describing well the beam wandering, broadening and deformation in weak and strong turbulence environment. In order to describe the effects of the atmospheric turbulence in more detail, the phase screen model [21] is taken advantage of modeling the effects of the atmosphere over the propagating beam, which presents lower variances in the simulation of the probability density function of the transmittance. In addition, maximum-likelihood estimation is widely used to estimate the channel parameters of the stabilization sub-channel by some variables in quantum signals, when the atmospheric channel for the free-space CV-QKD protocol can be deemed to the combination of several sub-channels with stability [22]. However, the parameter estimation for the stabilization sub-channel is required to sacrifice some variables to guarantee the estimation precision, which results in the cost of decreasing

This paper was supported by the National Natural Science Foundation of China (Grant No.11704412), Key Research and Development Program of Shaanxi (Program No. 2019ZDLGY09-01), Innovative Talents Promotion Plan in Shaanxi Province (Grant No. 2020KJXX-011) and Key Program of National University of Defense and Technology (19-QNCXJ-009). (Corresponding author: Xiaowen Liu and Chen Dong.)

X. W. Liu, C. Dong, and T. Y. Wu are with the School of Information and Communications, National University of Defense Technology, Xi'an, 710100, P. R. China. (e-mail: lxw5054@163.com; dongchengfkd@163.com; wtywzs@126.com;). X. Y. Wang is the Institute of Information and Navigation, Air Force Engineering University, Xi'an 710077, P. R. China. (e-mail: wang_kgd@foxmail.com).

the final secret key size [23]. It is urgent to estimate the channel parameters with as few variables as possible. The blind parameter estimation for CV-QKD over the atmospheric link has the capacity to calculating the estimated value of transmittance and excess with the higher-order statistics of each sub-channel rather than the variables [24], so that all the variables received by Bob can be used to generate the final secret key through reverse reconciliation and privacy amplification.

In this paper, a channel parameter estimation method based on compressive sensing (CS) theory is proposed for free-space CV-QKD, where the gaussian modulated coherent state is prepared with the modulation rate at the order of several MHz and the atmospheric channel is also deemed to several stabilization sub-channels. The application conditions of CS theory are introduced as well as the free-space CV-QKD protocol in the asymptotic regime at first, which is the foundation of adopting the CS theory to estimate the channel parameters and evaluating the practical security of the QKD protocol. Then after analyzing the possibility of the sparse representation for free-space channel, the parameter estimation methods based on CS are proposed according to the variables and their statistics, respectively. The sparse reconstruction model constructed with the variables of the quantum signals has a lower computational complexity due to no need to calculate the higher-order statistics and saves some variables for parameter estimation, while the sparse reconstruction model constructed with the second-order statistics of the variables only needs the variance of the variables to estimate the channel parameters without the need for the variables. Finally, the orthogonal matching pursuit (OMP) is applied to solve the two sparse reconstruction models, and the performance of the CS-based channel parameter estimation and the influence on free-space CV-QKD are analyzed by the simulation experiments which demonstrate that the proposed method can effectively calculate the channel parameters with high precision and low computational complexity and with need for fewer variables or no variables.

The organization of this paper is as follows. CS theory and free-space CV-QKD protocol are introduced in Section II. In Section III, the sparsity of the atmospheric channel in free-space CV-QKD is analyzed and two types of sparse reconstruction models for the channel parameters are constructed. The experiment results are shown in Section IV, while the conclusions are given in Section V.

## II. PRELIMINARIES

### A. CS theory

CS is a linear and non-adaptive sampling theory which is dependent of the sparsity of the data on a certain transform domain [25]. The technique is capable of breaking through the limitation of Nyquist sampling theorem and restoring the whole data by virtue of the partial measured value [26]. Thus, CS is widely used in the sparse data reconstructing in the underdetermined condition [27]. For the underdetermined equation as follows:

$$\mathbf{y} = \mathbf{\Phi}\mathbf{x} + \mathbf{z} = \mathbf{\Phi}\mathbf{\Psi}\mathbf{s} + \mathbf{z} = \mathbf{\Theta}\mathbf{s} + \mathbf{z} \qquad (1)$$

where $\mathbf{y}$ is a $N_s$-dimensional measured value, $\mathbf{\Phi} \in \mathbb{R}^{N_s \times N}$ is the measurement matrix with $N_s < N$, $\mathbf{x}$ is a $N$-dimensional original data, $\mathbf{s}$ which only has $K(K \ll N)$ nonzero elements is the sparse representation of the data $\mathbf{x}$ in the $\mathbf{\Psi}$ domain. The orthonormal transformation basis $\mathbf{\Psi} \in \mathbb{R}^{N \times N}$ is usually referred to as the sparse matrix which could be the inverse Fourier transform matrix, the wavelet transform and the discrete cosine transform, etc [16]. The $N_s \times N$ matrix $\mathbf{\Theta} = \mathbf{\Phi}\mathbf{\Psi}$ is called the sensing matrix. $\mathbf{z}$ is a $N$-dimensional additive noise.

When the measurement matrix $\mathbf{\Phi}$ satisfies the restricted isometry property (RIP) and the vector $\mathbf{s}$ only has $K(K \ll N)$ nonzero elements, the reconstruction methods such as the OMP algorithm and the basis pursuit (BP) algorithm can be applied to reconstruct the unknown vector $\mathbf{s}$ exactly from the underdetermined equation with the acquired vector $\mathbf{y}$ and matrix $\mathbf{\Theta}$ [28, 29]. The RIP provides the inequality constraint to decide the property of the measurement matrix $\mathbf{\Phi}$, which is given by

$$(1-\delta_K)\|\mathbf{x}\|_2^2 \leq \|\mathbf{\Phi}\mathbf{x}\|_2^2 \leq (1+\delta_K)\|\mathbf{x}\|_2^2 \qquad (2)$$

To simplify the operation complexity, the mutual incoherence property (MIP) which requires the incoherence between the different columns of the sensing matrix $\mathbf{\Theta}$ is proposed as an alternative criterion in the literature [30]. It is verified that the measurement matrix $\mathbf{\Phi}$ will satisfy the RIP with high probability if the MIP is achieved. The MIP can be expressed as

$$\mu(\mathbf{\Theta}) = \max_{1 \leq k \neq j \leq n} \left| \langle \theta_k, \theta_j \rangle \right| \qquad (3)$$

where $\theta_k$ and $\theta_j$ are two different columns of the sensing matrix $\mathbf{\Theta}$. There is stronger incoherence between the different columns of the matrix $\mathbf{\Theta}$ with smaller value of $\mu(\mathbf{\Theta})$. In addition, when the matrix $\mathbf{\Phi}$ is a random matrix such as the Gaussian matrix, Bernoulli matrix, and almost all other matrices with independent and identically distributed (i.i.d.) entries, the measurement matrix $\mathbf{\Phi}$ satisfies the RIP and the MIP with high probability and the matrix $\mathbf{A} = \mathbf{\Phi}\mathbf{H}$ where $\mathbf{H}$ is an arbitrary matrix also possesses the i.i.d. entries [31].

### B. System Model

Most CV-QKD systems employ a standard unified process. After Alice prepares and sends the random coherent states through an untrusted quantum channel, Bob measures the quantum state with the coherent detection and reconciliates the data with Alice. Then $m = N - n$ of the correlated data is publicly disclosed by Alice and Bob to estimate the parameters of the quantum channel and calculate the covariance matrix of the quantum states. Finally, the secret key is generated on both sides via the privacy amplification process in which a random hash function is adopted to two identical strings [32-34].

For the free-space continuous-variable quantum key distribution in which the quantum states propagate through the atmospheric channel, the transmittance and the excess noise are all fluctuating with time due to beam extinction and turbulence effects, etc. Thus, these channel parameters are random variables and the mean values of them in the actual propagation process need to be substituted into the covariance matrix of the quantum states to analyze the secret key rate and the effect of the fluctuating channel [23]. Considering that the modulation rate of the quantum state is at least three orders of magnitude higher than the fluctuation rate of the atmospheric channel, the atmospheric channel is capable of transmitting thousands of quantum states within the channel parameter change interval. Therefore, the whole atmospheric channel during the quantum state transmission can be divided into a number of sub-channels in which the channel parameters are consistent. The transmittance and the excess noise of the $i$-th sub-channel can be denoted by $T_i$ and $\varepsilon_i$, respectively. The mean value of the transmittance and the excess noise of the whole transmission channel be calculated as $\langle T \rangle = \sum_i^M p_i T_i$ and $\langle \varepsilon \rangle = \sum_i^M p_i \varepsilon_i$ where $M$ is the number of sub-channels and $p_i$ is the probability of $T_i$ and $\varepsilon_i$.

For the $i$-th sub-channel, the total additive noise contains the channel noise and the detection noise, and the variance of the total additive noise can be expressed as follows:

$$\begin{aligned}\chi_{\text{tot}} &= \chi_{\text{line}} + \frac{\chi_{\text{h}}}{T_i} \\ &= \left(\frac{1}{T_i} - 1 + \varepsilon_i\right) + \frac{(1-\eta) + \upsilon_{\text{el}}}{\eta T_i} \\ &= \frac{-\eta T_i + 1 + \eta T_i \varepsilon_i + \upsilon_{\text{el}}}{\eta T_i}\end{aligned} \quad (4)$$

where $\eta$ is the detection efficiency and $\upsilon_{\text{el}}$ is the electronic noise. The measured variance is given by

$$\begin{aligned}V_B &= \eta T_i (V_A + 1 + \chi_{\text{tot}}) \\ &= \eta T_i V_A + 1 + \eta T_i \varepsilon_i + \upsilon_{\text{el}}\end{aligned} \quad (5)$$

where $V_A$ is the modulation variance. The Alice's and Bob's variables for parameter estimation can be denoted by $\{x_j^i\}_{j=1,2,\ldots,m_i}^{i=1,2,\ldots,M}$ and $\{y_j^i\}_{j=1,2,\ldots,m_i}^{i=1,2,\ldots,M}$, respectively. $m_i$ is the number of the variables disclosed to estimate the parameters in the $i$-th sub-channel. The variables of Bob and Alice satisfy the equation as follows:

$$y_j^i = \sqrt{\eta T_i} x_j^i + z_j^i \quad (6)$$

where the mean and the variance of the noise $z_j^i$ are zero and $\sigma_i^2 = 1 + \eta T_i \varepsilon_i + \upsilon_{\text{el}}$.

### C. Secret Key Rate in Asymptotic Regime

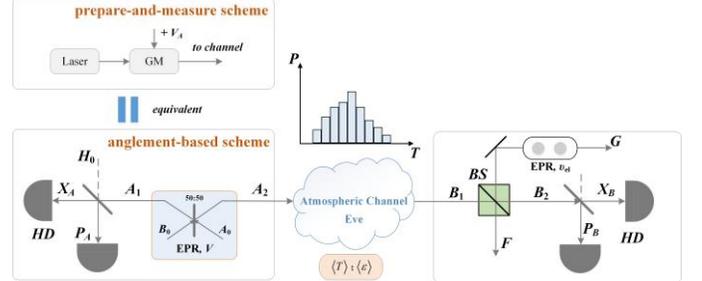

Fig. 1. Schematic diagram of free-space CVQKD. GM, BS and HD are the Gaussian modulation, the beam splitter and the homodyne detection, respectively.

The main work in this paper is to propose an effective channel parameter estimation method for free-space CV-QKD, while only the secret key rate in the asymptotic regime is calculated to analyze the influence of CS based parameter estimation method. The equivalent entanglement-based schematic diagram regarding the prepare-and-measure CV-QKD is depicted in Fig.1, where the EPR source with variance $V = V_A + 1$ produces an entangled state. One of the modes $A_2$ is sent to Bob through the fluctuating atmospheric channel and the mode $B_1$ be the mode received by Bob. Then Alice measures the preserved mode $A_1$, while Bob measures the received mode $B_1$ by using heterodyne detection to obtain the variables $x_{B_1}$ and $p_{B_1}$. Before the measurement of the mode $B_1$, the covariance matrix $\gamma_{A_1 B_1}$ is given by

$$\gamma_{A_1 B_1} = \begin{bmatrix} V \mathbb{1}_2 & \langle \sqrt{T} \rangle \sqrt{V^2 - 1} \sigma_z \\ \langle \sqrt{T} \rangle \sqrt{V^2 - 1} \sigma_z & \langle T \rangle \left( V + \frac{1}{\langle T \rangle} - 1 + \langle \varepsilon \rangle \right) \mathbb{1}_2 \end{bmatrix} \quad (7)$$

where $\mathbb{1}_2 = \text{diag}(1,1)$ is $2 \times 2$ unity matrix and $\sigma_z = \text{diag}(1,-1)$ is the Pauli matrix. $\langle T \rangle$, $\langle \sqrt{T} \rangle$ and $\langle \varepsilon \rangle$ are the mean of the random variable $T$, $\sqrt{T}$ and $\varepsilon$, respectively. The secret key rate $K$ in asymptotic regime and reverse reconciliation is given by

$$K = \beta I(A:B) - \chi(B:E) \quad (8)$$

where $\beta$ is the reconciliation efficiency, $I(A:B)$ is the mutual information extracted by Alice and Bob via the quantum state with the coherent detection and reconciliates the data with Alice. Then $m = N - n$ of the correlated data is publicly disclosed by Alice and Bob to estimate the parameters of the quantum channel and calculate the covariance matrix of the quantum states. Finally, the secret key is generated on both sides via the privacy amplification process in which a random hash function is adopted to two identical strings [32-34].

For the free-space continuous-variable quantum key distribution in which the quantum states propagate through the atmospheric channel, the transmittance and the excess noise are all fluctuating with time due to beam extinction and turbulence effects, etc. Thus, these channel parameters are random variables and the mean values of them in the actual propagation process need to be substituted into the covariance matrix of the quantum states to analyze the secret key rate and the effect of the fluctuating channel [23]. Considering that the modulation rate of the quantum state is at least three orders of magnitude higher than the fluctuation rate of the atmospheric channel, the atmospheric channel is capable of transmitting thousands of quantum states within the channel parameter change interval. Therefore, the whole atmospheric channel during the quantum state transmission can be divided into a number of sub-channels in which the channel parameters are consistent. The transmittance and the excess noise of the $i$-th sub-channel can be denoted by $T_i$ and $\varepsilon_i$, respectively. The mean value of the transmittance and the excess noise of the whole transmission channel be calculated as $\langle T \rangle = \sum_i^M p_i T_i$ and $\langle \varepsilon \rangle = \sum_i^M p_i \varepsilon_i$ where $M$ is the number of sub-channels and $p_i$ is the probability of $T_i$ and $\varepsilon_i$.

For the $i$-th sub-channel, the total additive noise contains the channel noise and the detection noise, and the variance of the total additive noise can be expressed as follows:

$$\begin{aligned}\chi_{\text{tot}} &= \chi_{\text{line}} + \frac{\chi_{\text{h}}}{T_i} \\ &= \left(\frac{1}{T_i} - 1 + \varepsilon_i\right) + \frac{(1-\eta) + \upsilon_{\text{el}}}{\eta T_i} \\ &= \frac{-\eta T_i + 1 + \eta T_i \varepsilon_i + \upsilon_{\text{el}}}{\eta T_i}\end{aligned} \quad (4)$$

where $\eta$ is the detection efficiency and $\upsilon_{\text{el}}$ is the electronic noise. The measured variance is given by

$$\begin{aligned}V_B &= \eta T_i (V_A + 1 + \chi_{\text{tot}}) \\ &= \eta T_i V_A + 1 + \eta T_i \varepsilon_i + \upsilon_{\text{el}}\end{aligned} \quad (5)$$

where $V_A$ is the modulation variance. The Alice's and Bob's variables for parameter estimation can be denoted by $\{x_j^i\}_{j=1,2,\ldots,m_i}^{i=1,2,\ldots,M}$ and $\{y_j^i\}_{j=1,2,\ldots,m_i}^{i=1,2,\ldots,M}$, respectively. $m_i$ is the number of the variables disclosed to estimate the parameters in the $i$-th sub-channel. The variables of Bob and Alice satisfy the equation as follows:

$$y_j^i = \sqrt{\eta T_i}\, x_j^i + z_j^i \quad (6)$$

where the mean and the variance of the noise $z_j^i$ are zero and $\sigma_i^2 = 1 + \eta T_i \varepsilon_i + \upsilon_{\text{el}}$.

### C. Secret Key Rate in Asymptotic Regime

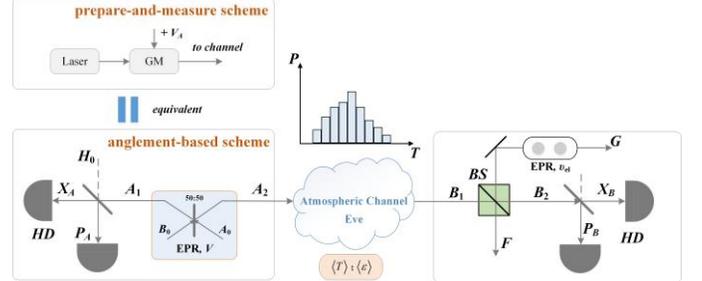

Fig. 1. Schematic diagram of free-space CVQKD. GM, BS and HD are the Gaussian modulation, the beam splitter and the homodyne detection, respectively.

The main work in this paper is to propose an effective channel parameter estimation method for free-space CV-QKD, while only the secret key rate in the asymptotic regime is calculated to analyze the influence of CS based parameter estimation method. The equivalent entanglement-based schematic diagram regarding the prepare-and-measure CV-QKD is depicted in Fig.1, where the EPR source with variance $V = V_A + 1$ produces an entangled state. One of the modes $A_2$ is sent to Bob through the fluctuating atmospheric channel and the mode $B_1$ be the mode received by Bob. Then Alice measures the preserved mode $A_1$, while Bob measures the received mode $B_1$ by using heterodyne detection to obtain the variables $x_{B_1}$ and $p_{B_1}$. Before the measurement of the mode $B_1$, the covariance matrix $\gamma_{A_1 B_1}$ is given by

$$\gamma_{A_1 B_1} = \begin{bmatrix} V \mathbb{1}_2 & \langle \sqrt{T} \rangle \sqrt{V^2 - 1}\, \sigma_z \\ \langle \sqrt{T} \rangle \sqrt{V^2 - 1}\, \sigma_z & \langle T \rangle \left( V + \frac{1}{\langle T \rangle} - 1 + \langle \varepsilon \rangle \right) \mathbb{1}_2 \end{bmatrix} \quad (7)$$

where $\mathbb{1}_2 = \text{diag}(1,1)$ is $2 \times 2$ unity matrix and $\sigma_z = \text{diag}(1,-1)$ is the Pauli matrix. $\langle T \rangle$, $\langle \sqrt{T} \rangle$ and $\langle \varepsilon \rangle$ are the mean of the random variable $T$, $\sqrt{T}$ and $\varepsilon$, respectively. The secret key rate $K$ in asymptotic regime and reverse reconciliation is given by

$$K = \beta I(A:B) - \chi(B:E) \quad (8)$$

where $\beta$ is the reconciliation efficiency, $I(A:B)$ is the mutual information extracted by Alice and Bob via the reconciliation and $\chi(B:E)$ is used to represent the Holevo information between Eve and Bob. In the consideration of the detection efficiency $\eta$ and electronic noise $\upsilon_{el}$, the mutual information between Alice and Bob can be further expressed as follows:

$$I(A:B) = \frac{1}{2}\log_2 \frac{V+1}{V - \left[\langle\sqrt{T}\rangle^2 (V^2-1)/\langle T\rangle(V+\chi_{tot})\right]+1}$$
$$= \frac{1}{2}\log_2 \frac{\langle T\rangle(V+\chi_{tot})}{\langle T\rangle(V+\chi_{tot}) - \langle\sqrt{T}\rangle^2 (V-1)} \quad (9)$$

The mathematical calculation of the Holevo information can be simplified to the equation as follows:

$$\chi(B:E) = \sum_{i=1}^{2} G(\lambda_i) + \sum_{i=3}^{5} G(\lambda_i) \quad (10)$$

where

$$G(x) = \frac{x+1}{2}\log_2\left(\frac{x+1}{2}\right) - \frac{x-1}{2}\log_2\left(\frac{x-1}{2}\right) \quad (11)$$

$\lambda_1$ and $\lambda_2$ are the symplectic eigenvalues of the covariance matrix $\gamma_{A_1 B_1}$, while $\lambda_3$, $\lambda_4$ and $\lambda_5$ are the symplectic eigenvalues of the covariance matrix $\gamma_{A_1 FG}^{m_B}$ where $m_B$ represents the measurement of Bob.

## III. CHANNEL PARAMETER ESTIMATION BASED ON CS

### A. Parameter Estimation by Variables

According to the relation between the variables of Bob and Alice (as shown in Eq. (6)), all Bob's variables through the $i$-th channel can be expressed with the vector as follows:

$$\mathbf{Y}^i = \mathbf{X}^i \mathbf{H}^i + \mathbf{Z}^i = \begin{pmatrix} y_1^i & y_2^i & \cdots & y_{m_i}^i \end{pmatrix}^T \quad (12)$$

where the matrix representation of Alice's variables, the transfer function and the noise can be defined, respectively, as follows:

$$\mathbf{X}^i = \begin{pmatrix} x_1^i & 0 & \cdots & 0 \\ 0 & x_2^i & \cdots & 0 \\ \vdots & \vdots & \ddots & \vdots \\ 0 & 0 & \cdots & x_{m_i}^i \end{pmatrix} \quad (13)$$

$$\mathbf{H}^i = \begin{pmatrix} \sqrt{\eta T_i} & \sqrt{\eta T_i} & \cdots & \sqrt{\eta T_i} \end{pmatrix}^T \quad (14)$$

$$\mathbf{Z}^i = \begin{pmatrix} z_1^i & z_2^i & \cdots & z_{m_i}^i \end{pmatrix}^T \quad (15)$$

Therefore, for the whole process of quantum state transmission, the vector representation of all Bob's variables is the combination of Bob's variables regarding each sub-channel, which can be expressed as follows:

$$\begin{pmatrix} \mathbf{Y}^1 \\ \mathbf{Y}^2 \\ \vdots \\ \mathbf{Y}^M \end{pmatrix} = \mathbf{Y} = \mathbf{XH} + \mathbf{Z} = \begin{pmatrix} \mathbf{X}^1 & & & \\ & \mathbf{X}^2 & & \\ & & \ddots & \\ & & & \mathbf{X}^M \end{pmatrix} \begin{pmatrix} \mathbf{H}^1 \\ \mathbf{H}^2 \\ \vdots \\ \mathbf{H}^M \end{pmatrix} + \begin{pmatrix} \mathbf{Z}^1 \\ \mathbf{Z}^2 \\ \vdots \\ \mathbf{Z}^M \end{pmatrix} \quad (16)$$

Considering that Alice's data $\{x_j^i\}_{j=1,2,\ldots,m_i}^{i=1,2,\ldots,M}$ are Gaussian random variables with the mean of 0 and the variance of $V_A$, it can be inferred that the matrix $\mathbf{X}$ and $\mathbf{X}^i$ are Gaussian matrix and satisfy the RIP and the MIP with high probability. Moreover, although the matrix $\mathbf{H}$ and $\mathbf{H}^i$ of which all the elements are almost nonzero are the non-sparse matrixes, all the elements of the matrix $\mathbf{H}^i$ are identical and the matrix $\mathbf{H}$ only has $M$ different kinds of value, so that the matrix $\mathbf{H}$ and $\mathbf{H}^i$ have the potential of being represented sparsely. It is generally known that the Fourier transform of the rectangular function is sinc function. Especially, when the width of the rectangular function tends to infinity, sinc function will turn into impulse function. In sinc function and impulse function, the number of nonzero elements is far more than that of zero elements. Therefore, the matrix of sinc function or impulse function can be regarded as the sparse representation of the matrix $\mathbf{H}$ and $\mathbf{H}^i$ in the discrete Fourier transform (DFT) domain. Let $\mathbf{S}$ and $\mathbf{S}^i$ be the sparse representation of the matrix $\mathbf{H}$ and $\mathbf{H}^i$ in the DFT domain, respectively. The matrix $\mathbf{S}$ (or $\mathbf{S}^i$) and the matrix $\mathbf{H}$ (or $\mathbf{H}^i$) satisfy the equation as follows:

$$\mathbf{H} = \mathbf{\Psi S} = \mathbf{D}^{-1}\mathbf{S} \quad (17)$$

$$\mathbf{H}^i = \mathbf{\Psi}^i \mathbf{S}^i = \mathbf{D}_i^{-1}\mathbf{S}^i \quad (18)$$

where $\mathbf{D}_i^{-1}$ is a $m_i \times m_i$ inverse DFT (IDFT) matrix for the $i$-th sub-channel, while $\mathbf{D}^{-1}$ is a $m \times m$ IDFT matrix for the whole transmission process. Therefore, all Bob's variables through the whole channel and the $i$-th sub-channel can be rewritten, respectively, as follows:

$$\mathbf{Y} = \mathbf{X\Psi S} + \mathbf{Z} \quad (19)$$

$$\mathbf{Y}^i = \mathbf{X}^i \mathbf{\Psi}^i \mathbf{S}^i + \mathbf{Z}^i \quad (20)$$

As analyzed above, $\mathbf{X}$ and $\mathbf{X}^i$ are the random matrix and satisfy the RIP and the MIP with high probability, and $\mathbf{S}$ and $\mathbf{S}^i$ only have very few nonzero element. The unknown vector $\mathbf{S}$ and $\mathbf{S}^i$ can be reconstructed from (19) and (20) by the CS reconstruction methods, even if the variables of Alice and Bob are abstracted by sparse sampling. Let $m_s^i$ ($m_s^i < m_i$) be the number of the variables of Alice and Bob by sparse sampling regarding the $i$-th sub-channel. When only using these variables to estimate the channel parameters, more quantum signals are saved to produce the secret key and it have a negligible effect on parameter estimation accuracy. The $m_s^i \times m_i$ matrix $\mathbf{\Phi}^i$ for sparse sampling can be defined as follows:

$$\Phi_{j,k}^i = \begin{cases} 1, & \{(j,k)|k = \varphi_j^i\} \\ 0, & \text{others} \end{cases} \quad (21)$$

where $\varphi_j^i$ is the element in the vector $\boldsymbol{\varphi} = \{\varphi_1^i, \ldots, \varphi_j^i, \ldots, \varphi_{m_s^i}^i\}$ which is constituted by $m_s^i$ different integers chosen at random from 0 to $m_i$. The sparse sampling matrix regarding the whole transmission process can be defined as follows:

$$\mathbf{\Phi} = \begin{pmatrix} \mathbf{\Phi}^1 & & & \\ & \mathbf{\Phi}^2 & & \\ & & \ddots & \\ & & & \mathbf{\Phi}^M \end{pmatrix} \quad (22)$$

In the situation of sparse sampling, the sparse measurement values regarding the whole channel and the $i$-th sub-channel can be rewritten, respectively, as follows:

$$\mathbf{Y}_s = \mathbf{X}_s \mathbf{\Psi} \mathbf{S} + \mathbf{Z}_s \quad (23)$$

$$\mathbf{Y}_s^i = \mathbf{X}_s^i \mathbf{\Psi}^i \mathbf{S}^i + \mathbf{Z}_s^i \quad (24)$$

where the $\sum m_s^i \times 1$ vector $\mathbf{Y}_s = \mathbf{\Phi}\mathbf{Y}$, the $\sum m_s^i \times \sum m_i$ matrix $\mathbf{X}_s = \mathbf{\Phi}\mathbf{X}$, the $\sum m_s^i \times 1$ vector $\mathbf{Z}_s = \mathbf{\Phi}\mathbf{Z}$, the $m_s^i \times 1$ vector $\mathbf{Y}_s^i = \mathbf{\Phi}^i \mathbf{Y}^i$, the $m_s^i \times m_i$ matrix $\mathbf{X}_s^i = \mathbf{\Phi}^i \mathbf{X}^i$, the $m_s^i \times 1$ vector $\mathbf{Z}_s^i = \mathbf{\Phi}^i \mathbf{Z}^i$. For (23) and (24), they are the underdetermined equation and the matrix $\mathbf{X}_s \mathbf{\Psi}$ and $\mathbf{X}_s^i \mathbf{\Psi}^i$ also possesses the i.i.d. entries. Therefore, even if the sparse variable is used to estimate the channel parameters, the vector $\mathbf{S}$ (or $\mathbf{S}^i$) which is the DFT of the matrix $\mathbf{H}$ (or $\mathbf{H}^i$) and only has $K$ ($K \ll m_i$) nonzero elements can be reconstructed by solving the sparse reconstruction model. Taking the $i$-th sub-channel as example, the sparse reconstruction model can be defined as follows:

$$\hat{\mathbf{S}}^i = \min \left\| \mathbf{\Psi}^{i\mathrm{H}} \mathbf{X}_s^{i\mathrm{H}} \mathbf{Y}_s^i \right\|_1 \\ \text{s.t.} \left\| \mathbf{Y}_s^i - \mathbf{X}_s^i \mathbf{\Psi}^i \hat{\mathbf{S}}^i \right\|_2 \leq \delta \quad (25)$$

For the typical CS reconstruction problem, the OMP algorithm can be adopted to reconstruct the vector $\mathbf{S}^i$. Let $\hat{\mathbf{S}}^i$ be the reconstructed vector $\mathbf{S}^i$. Then the estimated matrix $\mathbf{H}^i$ can be calculated as follows:

$$\hat{\mathbf{H}}^i = \mathbf{\Psi}^i \hat{\mathbf{S}}^i = \mathbf{D}_i^{-1} \hat{\mathbf{S}}^i \quad (26)$$

Therefore, the estimated transmittance of the $i$-th sub-channel can be calculated as follows:

$$\hat{T}_i = \frac{\left( \sum \hat{\mathbf{H}}^i \right)^2}{\eta \cdot m_i^2} \quad (27)$$

According to the relation among the transmittance, the excess noise and the variance of Alice's and Bob's variables, the estimated excess noise regarding the $i$-th sub-channel can be calculated as follows:

$$\hat{\varepsilon}_i = \frac{\mathbf{Y}_s^{i\mathrm{H}} \mathbf{Y}_s^i - \eta \hat{T}_i \sum \mathbf{X}_s^i \mathbf{X}_s^{i\mathrm{H}} - m_s^i (1 + \upsilon_{\mathrm{el}})}{m_s^i \eta \hat{T}_i} \quad (28)$$

It is worth noting that the sufficiently small noise $\mathbf{Z}_s^i$ is also an important factor to guarantee that the vector $\mathbf{S}^i$ in (24) can be reconstructed by solving the sparse reconstruction model (as shown in (25)). For free-space CV-QKD, the noise $\mathbf{Z}_s^i$ is a random variable with the mean of 0 and the variance of $\sigma_i^2 = 1 + \eta T_i \varepsilon_i + \upsilon_{\mathrm{el}}$. Thereinto, $\eta T_i \varepsilon_i$ and $\upsilon_{\mathrm{el}}$ are so small that the noise $\mathbf{Z}_s^i$ is small enough to reconstruct the vector $\mathbf{S}^i$ by CS method.

### B. Parameter Estimating by Statistics

When the variables of Alice and Bob are used to estimate the channel parameters, the number of the data used to generate the secret key will be reduced. Although the theory of CS provides the fundamental for a lower number of the required variables in the condition of guaranteeing the precise of the parameter estimation, there are still some variables which have to be publicly disclosed and sacrificed. In order to further reduce the variables for the parameter estimation, it is important to consider how to estimate the channel parameters by using the statistics of the variables of Alice and Bob. If only the received quantum states and the statistics of the transmitted states as the prior knowledge are applied to estimate the channel parameters, there would be no need to publicly disclose some data and the secret key rate of free-space CV-QKD would be improved.

Considering of the relation between the measured variance and the modulation variance as shown in (5), no data need to be sacrificed for the parameter estimation when the modulation variance is disclosed by Alice. For the $i$-th sub-channel, the $m_i \times 1$ vector of the measured variance can be expressed as follows:

$$\mathbf{R}_y^i = \mathbf{R}_x^i \mathbf{H}_R^i + \mathbf{R}_z^i = \begin{pmatrix} V_B^i & V_B^i & \cdots & V_B^i \end{pmatrix}^T \quad (29)$$

where $\mathbf{R}_x^i \in \mathbb{R}^{m_i \times m_i}$, $\mathbf{H}_R^i \in \mathbb{R}^{m_i}$ and $\mathbf{R}_z^i \in \mathbb{R}^{m_i}$ are composed of the modulation variance, the transmission and the noise variance regarding the $i$-th sub-channel, which can be expressed as follows:

$$\mathbf{R}_x^i = \begin{pmatrix} V_A^i & 0 & \cdots & 0 \\ 0 & V_A^i & \cdots & 0 \\ \vdots & \vdots & \ddots & \vdots \\ 0 & 0 & \cdots & V_A^i \end{pmatrix} \quad (30)$$

$$\mathbf{H}_R^i = \begin{pmatrix} \eta T_i & \eta T_i & \cdots & \eta T_i \end{pmatrix}^T \quad (31)$$

$$\mathbf{R}_z^i = \begin{pmatrix} \sigma_i^2 & \sigma_i^2 & \cdots & \sigma_i^2 \end{pmatrix}^T \quad (32)$$

Similarly, for the whole process of quantum state transmission, the vector representation of the measured variance is the combination of measured variance regarding each sub-channel, which can be expressed as follows:

$$\mathbf{R}_y = \mathbf{R}_x \mathbf{H}_R + \mathbf{R}_z = \begin{pmatrix} \mathbf{R}_y^1 \\ \mathbf{R}_y^2 \\ \vdots \\ \mathbf{R}_y^M \end{pmatrix}$$
$$= \begin{pmatrix} \mathbf{R}_x^1 & & & \\ & \mathbf{R}_x^2 & & \\ & & \ddots & \\ & & & \mathbf{R}_x^M \end{pmatrix} \begin{pmatrix} \mathbf{H}_R^1 \\ \mathbf{H}_R^2 \\ \vdots \\ \mathbf{H}_R^M \end{pmatrix} + \begin{pmatrix} \mathbf{R}_z^1 \\ \mathbf{R}_z^2 \\ \vdots \\ \mathbf{R}_z^M \end{pmatrix} \quad (33)$$

On account of the identity of the elements in the vector $\mathbf{H}_R^i$ and $\mathbf{H}_R$, the two vectors possess the corresponding sparse representation in the DFT domain so that the two vectors can be further represented by the sparse representation $\mathbf{S}_R^i$ and $\mathbf{S}_R$ with the IDFT matrix $\mathbf{D}_i^{-1}$ and $\mathbf{D}^{-1}$. In addition, It is also feasible to reduce the data volume of the vector $\mathbf{R}_y^i$ (or $\mathbf{R}_y$) and the matrix $\mathbf{R}_x^i$ (or $\mathbf{R}_x$) by using the matrix $\mathbf{\Phi}^i$ and $\mathbf{\Phi}$ (as shown in (22)), respectively. Therefore, after the sparse representation and the random sparse sampling, the measured variance can be rewritten as follows:

$$\mathbf{R}_{s,y}^i = \mathbf{\Phi}^i \mathbf{R}_x^i \mathbf{D}_i^{-1} \mathbf{S}_R^i + \mathbf{\Phi}^i \mathbf{R}_z^i$$
$$= \mathbf{\Phi}^i \mathbf{R}_x^i \mathbf{\Psi}^i \mathbf{S}_R^i + \mathbf{\Phi}^i \mathbf{R}_z^i = \mathbf{\Theta}_R^i \mathbf{S}_R^i + \mathbf{\Phi}^i \mathbf{R}_z^i \quad (34)$$

$$\mathbf{R}_{s,y} = \mathbf{\Phi} \mathbf{R}_x \mathbf{D}^{-1} \mathbf{S}_R + \mathbf{\Phi} \mathbf{R}_z$$
$$= \mathbf{\Phi} \mathbf{R}_x \mathbf{\Psi} \mathbf{S}_R + \mathbf{\Phi} \mathbf{R}_z = \mathbf{\Theta}_R \mathbf{S}_R + \mathbf{\Phi} \mathbf{R}_z \quad (35)$$

where the $m_s^i \times 1$ vector $\mathbf{R}_{s,y}^i = \mathbf{\Phi}^i \mathbf{R}_y^i$ and the $\sum m_s^i \times 1$ vector $\mathbf{R}_{s,y} = \mathbf{\Phi} \mathbf{R}_y$. The matrix $\mathbf{\Theta}_R^i$ and $\mathbf{\Theta}_R$ possess the i.i.d. entries and satisfy the RIP and the MPC with high probability. The unknown vector $\mathbf{S}_R^i$ and $\mathbf{S}_R$ only has $K_i (K_i \ll m_i)$ and $K(K \ll \sum m_i)$ nonzero elements, respectively. However, the elements in the vector $\mathbf{\Phi}^i \mathbf{R}_z^i$ regarding the $i$-th sub-channel is $\sigma_i^2 = 1 + \eta T_i \varepsilon_i + \upsilon_{el}$ which is not small enough to reconstruct the unknown vector $\mathbf{S}_R^i$ from (34) by the CS method. Considering that $1 + \upsilon_{el}$ in $\sigma_i^2$ is almost seen as a constant, the $1 + \upsilon_{el}$ term can be separated from $\sigma_i^2$ in order to reduce the value of the noise term in (34), which can be expressed as follows:

$$\mathbf{R}_z^i = \mathbf{R}_T^i + \mathbf{R}_\upsilon^i \quad (36)$$

where

$$\mathbf{R}_T^i = \begin{pmatrix} \eta T_i \varepsilon_i & \eta T_i \varepsilon_i & \cdots & \eta T_i \varepsilon_i \end{pmatrix}^T \quad (37)$$

$$\mathbf{R}_\upsilon^i = \begin{pmatrix} 1 + \upsilon_{el} & 1 + \upsilon_{el} & \cdots & 1 + \upsilon_{el} \end{pmatrix}^T \quad (38)$$

Substituting (37) and (38) into (34), (34) can be rewritten as follows:

$$\mathbf{R}_{s,y}^i = \mathbf{\Theta}_R^i \mathbf{S}_R^i + \mathbf{\Phi}^i (\mathbf{R}_\varepsilon^i + \mathbf{R}_\upsilon^i) \quad (39)$$

(39) can be further rewritten as follows:

$$\mathbf{R}_{\upsilon,y}^i = \mathbf{\Theta}_R^i \mathbf{S}_R^i + \mathbf{R}_{s,\varepsilon}^i \quad (40)$$

where

$$\mathbf{R}_{\upsilon,y}^i = \mathbf{\Phi}^i (\mathbf{R}_y^i - \mathbf{R}_\upsilon^i) \quad (41)$$

$$\mathbf{R}_{s,\varepsilon}^i = \mathbf{\Phi}^i \mathbf{R}_\varepsilon^i \quad (42)$$

For (40), the value $\eta T_i \varepsilon_i$ in $\mathbf{R}_{s,\varepsilon}^i$ is small enough to reconstruct the vector $\mathbf{S}_R^i$ by the CS method. Similarly, for the whole transmission process, (35) can also be turn to the form which meet the requirement of CS theory and can be expressed as follows:

$$\mathbf{R}_{\upsilon,y} = \mathbf{\Theta}_R \mathbf{S}_R + \mathbf{R}_{s,\varepsilon} \quad (43)$$

where

$$\mathbf{R}_{\upsilon,y} = \mathbf{\Phi}(\mathbf{R}_y - \mathbf{R}_\upsilon) \quad (44)$$

$$\mathbf{R}_{s,\varepsilon} = \mathbf{\Phi}\mathbf{R}_{\varepsilon} = \mathbf{\Phi}\begin{bmatrix} \mathbf{R}_{\varepsilon}^{1} & \mathbf{R}_{\varepsilon}^{2} & \cdots & \mathbf{R}_{\varepsilon}^{M} \end{bmatrix}^{\mathrm{T}} \quad (45)$$

For (40) and (43), the corresponding sparse reconstruction model can be constructed. Taking the *i*-th sub-channel as example, the sparse reconstruction model regarding (40) can be defined as follows:

$$\begin{aligned} \hat{\mathbf{S}}_{\mathrm{R}}^{i} &= \min \left\| \mathbf{\Theta}_{\mathrm{R}}^{i\,\mathrm{H}} \mathbf{R}_{\upsilon,y}^{i} \right\|_{1} \\ \text{s.t.} \quad & \left\| \mathbf{R}_{\upsilon,y}^{i} - \mathbf{\Theta}_{\mathrm{R}}^{i} \hat{\mathbf{S}}_{\mathrm{R}}^{i} \right\|_{2} \leq \delta \end{aligned} \quad (46)$$

where $\hat{\mathbf{S}}_{\mathrm{R}}^{i}$ is the reconstructed vector $\mathbf{S}_{\mathrm{R}}^{i}$, and it can be obtained by adopting the OMP algorithm to solve (46). Then the estimated vector $\mathbf{H}_{\mathrm{R}}^{i}$ can be calculated as follows:

$$\hat{\mathbf{H}}_{\mathrm{R}}^{i} = \mathbf{\Psi}^{i} \hat{\mathbf{S}}_{\mathrm{R}}^{i} = \mathbf{D}_{i}^{-1} \hat{\mathbf{S}}_{\mathrm{R}}^{i} \quad (47)$$

Therefore, the estimated transmittance of the *i*-th sub-channel can be calculated as follows:

$$\hat{T}_{i} = \frac{\sum \hat{\mathbf{H}}_{\mathrm{R}}^{i}}{\eta \cdot m_{i}} \quad (48)$$

According to the relation among the transmittance, the excess noise, the modulation variance, and the measured variance, the estimated excess noise regarding the *i*-th sub-channel can be calculated as follows:

$$\hat{\varepsilon}_{i} = \frac{\sum \mathbf{R}_{\upsilon,y}^{i} - \eta \hat{T}_{i} \sum \mathbf{R}_{s,x}^{i}}{m_{s}^{i} \eta \hat{T}_{i}} \quad (49)$$

For the whole transmission process, the estimated transmittance and excess noise can be obtained by solving the corresponding sparse reconstruction model and also can be obtained by calculating the mean of the estimated value of the sub-channel, which be given by

$$\langle \hat{T} \rangle = \sum_{i=1}^{M} \hat{T}_{i} p_{i} \quad (50)$$

$$\langle \hat{\varepsilon} \rangle = \sum_{i=1}^{M} \hat{\varepsilon}_{i} p_{i} \quad (51)$$

*C. Performance Analysis*

Mean square error (MSE) is often used to evaluate the performance of the parameter estimation. For the proposed CS based channel parameter estimation for free-space CV-QKD, the MSE can be calculated as follows:

$$\mathrm{MSE} = \frac{1}{M} \sum_{i=1}^{M} \left( \hat{T}_{i} - T_{i} \right)^{2} \quad (52)$$

where $\hat{T}_{i}$ and $T_{i}$ are the estimated value and practical value of the transmittance of the *i*-th sub-channel, respectively.

## IV. EXPERIMENTS

The simulation experiments are given to analyze the performance of the CS-based channel parameter estimation method and the influence of the CS-based estimation on the secret key rate of free-space CV-QKD.

*A. Channel Parameter Estimation*

In the experiments, at first 100 sub-channels with the transmittance $T_{1}, T_{2}, \ldots, T_{100}$ which are obtained by the elliptical beam model are selected to form the integrated channel of the whole transmission process and suppose excess noise regarding each sub-channel is $\varepsilon$. Taking the 20 km propagation distance as example, let *N* which is the number of the variables for the parameter estimation be $10^{6}$. 10% and 100% are selected as the quantity of the random sparse sampling, thus $N/10$ and $N$ variables are used to estimate the channel parameters in practice, respectively. According to (24) and (40), the sensing matrixes in the sparse reconstruction model in virtue of variables and statistics are constructed and used to assess the reconfigurability by calculating the corresponding MIP, respectively. The calculation results of MIP regarding 100 sub-channels are recorded and illustrated as follows:

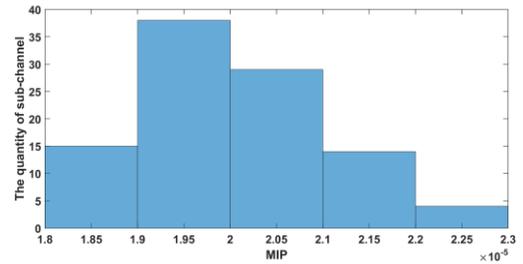

(a)

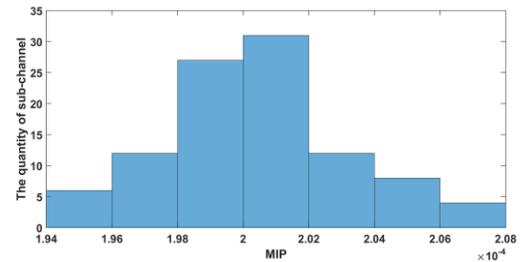

(b)

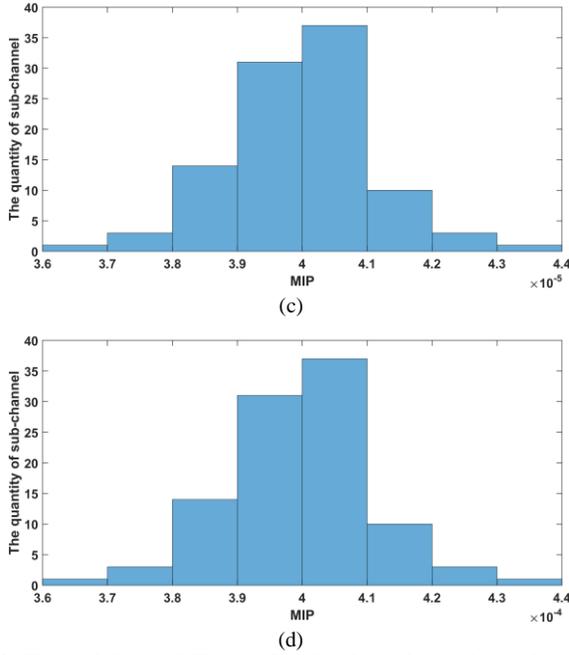

(c)

(d)

Fig. 2. The statistics on MIP. (a) 10% data is used to estimate the channel parameters in the model constructed by variables. (b) all the data is used to estimate the channel parameters in the model constructed by variables. (c) 10% data is used to estimate the channel parameters in the model constructed by statistics. (d) all data is used to estimate the channel parameters in the model constructed by statistics.

When Alice's and Bob's variables are used to constructed the sensing matrix of the sparse reconstruction model, the statistics of the MIP is depicted in Fig. 2 (a) and (b) where the maximum MIP is not more than $2.3 \times 10^{-5}$ and $2.08 \times 10^{-4}$ in two sparse cases. The MIP of most sub-channels are concentrated at approximately $1.95 \times 10^{-5}$ and $2.01 \times 10^{-4}$, respectively. While the statistics are used to constructed the sensing matrix of the sparse reconstruction model, the statistics of the MIP are depicted in Fig. 2 (c) and (d) where the maximum MIP is not more than $4.4 \times 10^{-5}$ and $4.4 \times 10^{-4}$ in two sparse cases. The MIP of most sub-channels is also concentrated at approximately $4.05 \times 10^{-5}$ and $4.05 \times 10^{-4}$, respectively. It is observed that the MIPs in two different models are so small that the vector $\mathbf{S}^i$ and $\mathbf{S}_R^i$ may be reconstructed by OMP algorithm with high probability.

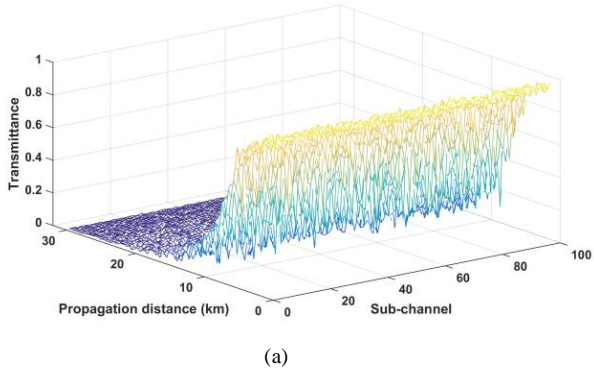

(a)

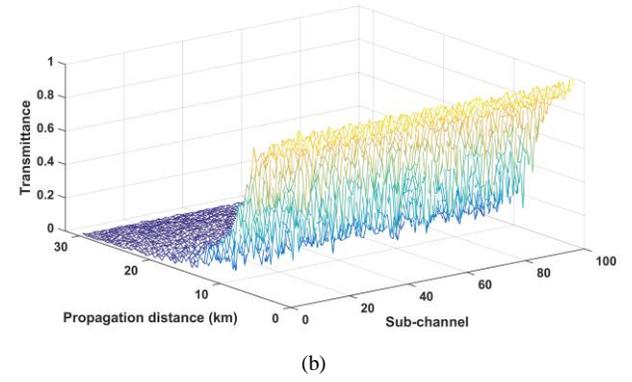

(b)

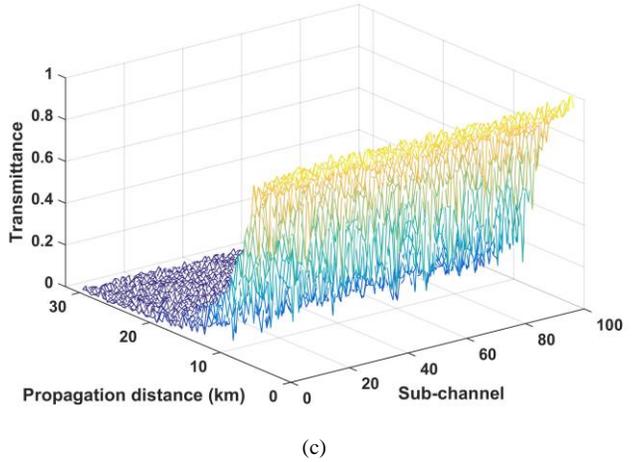

(c)

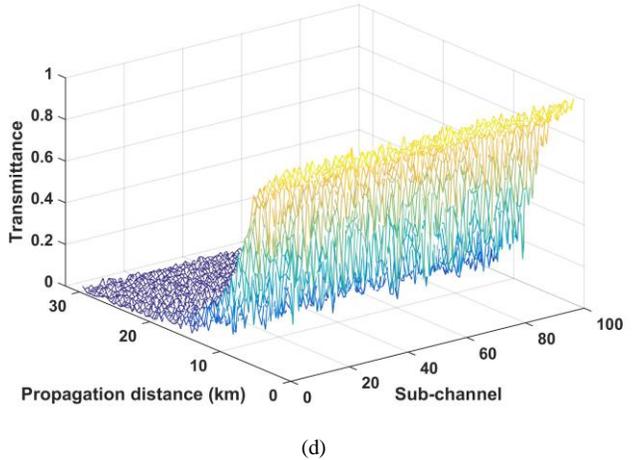

(d)

Fig. 3. The estimated transmittance regarding propagation distance and sub-channel. (a) 100% variables are used to estimate the channel parameters. (b) 40% variables are used to estimate the channel parameters. (c) 100% statistics are used to estimate the channel parameters. (d) 40% statistics are used to estimate the channel parameters.

Depending on the Alice's and Bob's variables and the sparse reconstruction model (as shown in (25)), the estimated transmittance regarding each sub-channel is reconstructed by solving (25) and (46) with OMP algorithm, respectively, as illustrated in Fig. 3 where the estimated results in the sparsity of 100% and 40% are depicted. It is observed that the estimated transmittance become so small over 14 km that scarcely any secret key may be shared. The estimation error can be quantitatively expressed according to (52) and is illustrated in Fig. 4 where the intensity value of the image pixel is the MSE. The mean values of the MSE of all the propagation distance for different sparsity are depicted with the red number in the middle. It is observed that the MSE does not change much even though the sparsity of the variable used to estimation reaches 40%. In comparison, the sparsity of 1% have almost no influence on the MSE when the statistics are used to estimate the channel parameters. Therefore, the compressive sensing based parameter estimation via the statistics not only can sacrifice no variables in the quantum signal, but also can reduce the computational complexity to a greater extent. However, the CS-based transmittance reconstruction method by using the variables has a higher estimation precision in comparison with that by the statistics when the sparsity is greater than 40%.

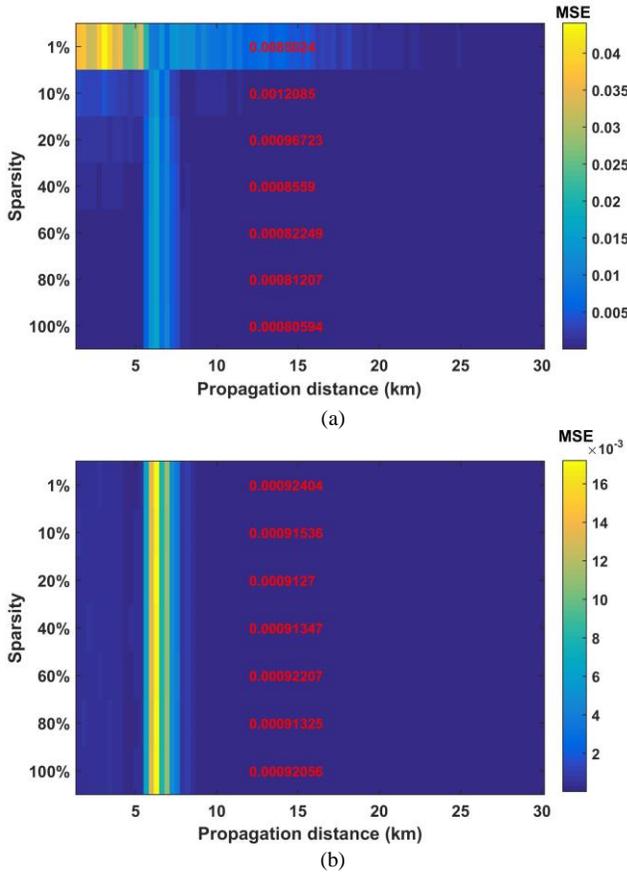

Fig. 4. The MSE of parameter estimation. The intensity value of the image pixel is the MSE, and the red numbers in the middle represent the mean values of the MSE of all the propagation distance for different sparsity. (a) the variables are used to estimate the channel parameters. (b) the statistics are used to estimate the channel parameters.

### B. Secret Key Rate

At first the results of the channel parameter estimation regarding 100 sub-channels with $10^6$ variables are substituted into the calculation formulas of the secret key rate to analyze the influence of the parameter estimation method. The quantity of the random sparse sampling be selected as 100%, and the secret key rate regarding different propagation distances are illustrated as Fig. 5 (a) and Fig. 6 (a) where the secret key rate with homodyne detection is higher than the secret key rate with heterodyne detection in short propagation distance. This is because the heterodyne detection will introduce more electronic noise. In addition, for the two detection means, it can be seen that the secret key rates obtained by using the estimated transmittance and the practical transmittance are very similar, especially when the propagation distance is no more than 11km. As shown in Fig. 5 (b) and Fig. 6 (b), several cases about the sparsity is considered, and the corresponding transmittance estimated by the variables and the statistics, respectively, is used to calculate the secret key rate. By comparing the secret key rate under the different sparsity, we can draw the conclusion that the secret key rate estimation precision is decreased with reducing the quantity of the random sparse sampling. Because the probability density functions of the transmittance at different distances overlap, especially at long distances, the sampling point of the transmittance by the elliptical beam model may appear the situation that it is possible to obtain a bigger transmittance for a longer distance. Thus, the secret key rate regarding the long propagation distance might be bigger than that regarding a little closer distance, such as more than 10 km in Fig. 5 (a)-(b) and Fig. 6 (a)-(b).

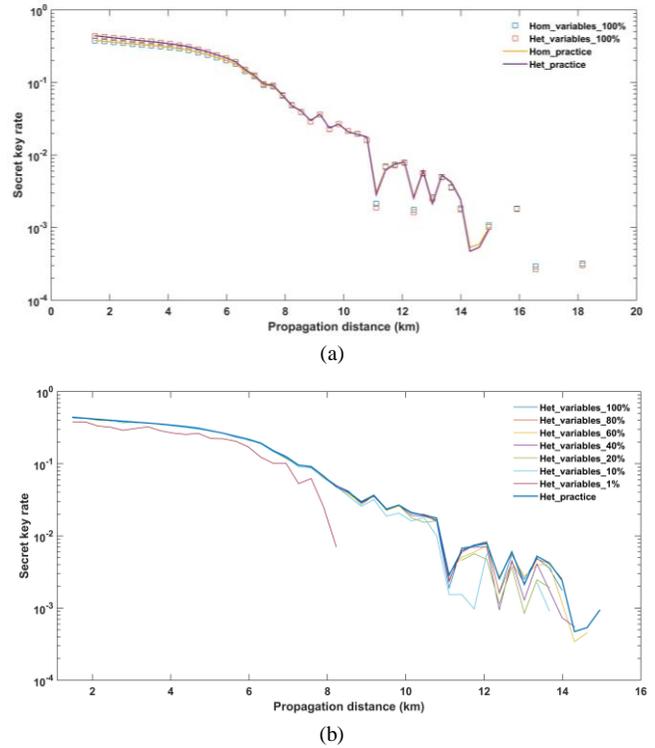

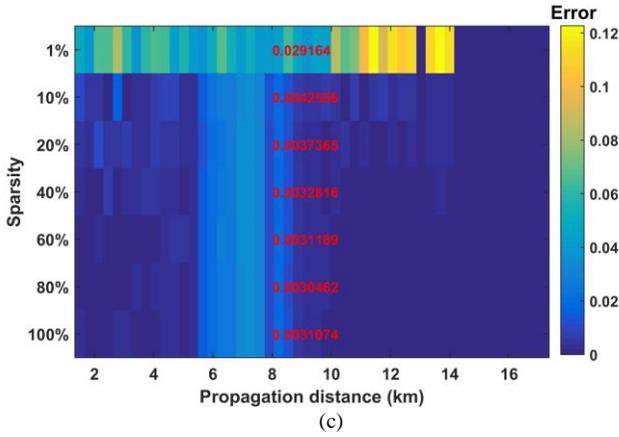

(c)

Fig. 5. The analysis on the secret key rate with the channel parameters estimated by the variables. (a) the secret key rate with the different detection means. (b) the secret key rate with the different sparsity. (c) The absolution error of secret key rate between the estimated parameter and the practice parameter, where the intensity value of the image pixel is the absolution error and the red numbers in the middle represent the mean values of the error of all the propagation distance for different sparsity.

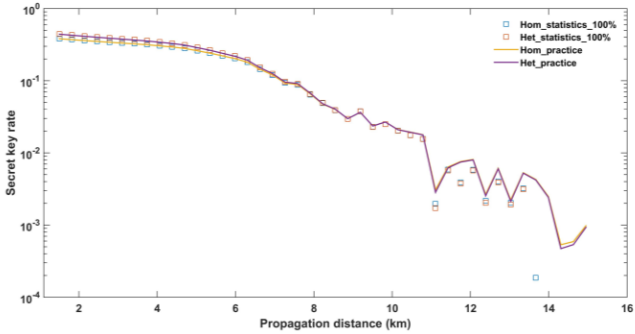

(a)

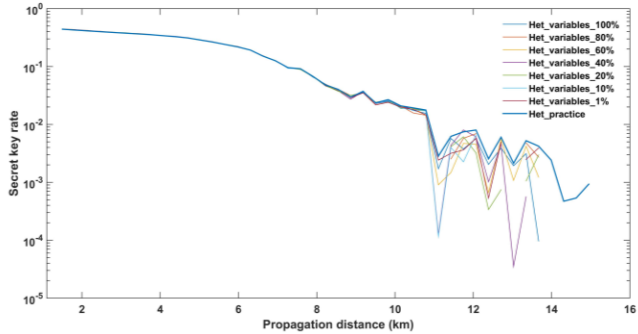

(b)

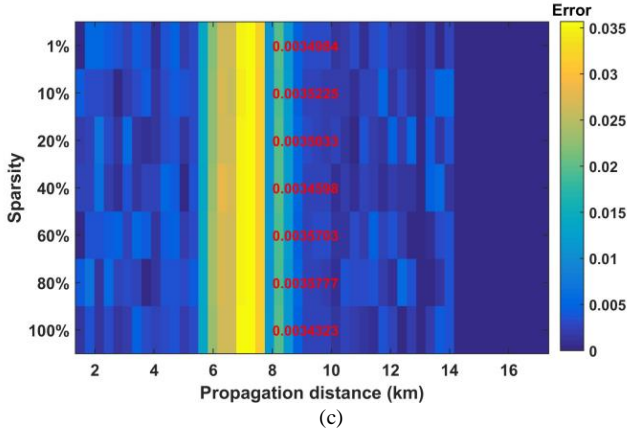

(c)

Fig. 6. The analysis on the secret key rate with the channel parameters estimated by the statistics. (a) the secret key rate with the different detection means. (b) the secret key rate with the different sparsity. (c) The absolution error of secret key rate between the estimated parameter and the practice parameter, where the intensity value of the image pixel is the absolution error and the red numbers in the middle represent the mean values of the error of all the propagation distance for different sparsity.

In summary, it can be concluded that the influence of the estimated channel parameters on the secret key rate is very small, especially when the statistics of the transmitted states are used to estimate the channel parameters, not only all the variables are used to generate the secret key to maximize the secret key, but also the quantity of the random sparse sampling can be reduced by about 100 times to reduce the computational complexity significantly. When the variables in quantum signals are used to estimate the channel parameters, the sparsity above 40% can remain a higher estimation precision in the channel parameters and the secret key rate so that about 60% variables used to parameter estimation can be saved to generate the secret key. In addition, it is observed that the secret key rate error between the estimated transmittance by statistics and practical transmittance is more apparent at the propagation distance of more than 10 km, while the secret key rate calculated by the estimated transmittance by variables is more similar to that by the practical transmittance. The result is consistent to the conclusion that the CS-based transmittance reconstruction method by using the variables has a higher the estimation precision with high quantity of the random sparse sampling.

## V. CONCLUSIONS

In this paper, the sparse reconstruction models for the channel parameters has been constructed by the variables and statistics, respectively, after analyzing the sparsity of the atmospheric channel in free-space CV-QKD protocol. OMP algorithm is adopted to calculate the estimated value of the parameters by solving the sparse reconstruction models and the numerical methods is executed to simulate the performance of the proposed method. The simulation results demonstrate that the MIPs are small enough to verify the sparsity of the atmospheric channel. When eighty percent of variables is saved from the parameter estimation, the MSE is also at the order of magnitude of $10^{-4}$. While one percent of the amount of the statistics is used to construct the sparse reconstruction model, low MSE can also be realized. Therefore, the proposed methods which are well adapted for the cases with the limited communication time not only reduce the variables for parameter estimation, even without the need for variables, but also decrease the computational complexity. The calculated MSE and the comparison of secret key rate indicate the high estimation precision of the proposed method.

## ACKNOWLEDGMENTS

This paper was supported by the National Natural Science Foundation of China (Grant No.11704412), Key Research and Development Program of Shaanxi (Program No.

2019ZDLGY09-01), Innovative Talents Promotion Plan in Shaanxi Province (Grant No. 2020KJXX-011) and Key Program of National University of Defense and Technology (19-QNCXJ-009).